\documentclass{article}
\usepackage{arxiv}

\usepackage{graphicx}%
\usepackage{multirow}%
\usepackage{amsmath,amssymb,amsfonts}%
\usepackage{amsthm}%
\usepackage{mathrsfs}%
\usepackage[title]{appendix}%
\usepackage{xcolor}%
\usepackage{textcomp}%
\usepackage{manyfoot}%
\usepackage{hyperref}
\usepackage{natbib}
\usepackage{threeparttable}
\usepackage{booktabs}%
\usepackage{algorithm}%
\usepackage{algorithmicx}%
\usepackage{algpseudocode}%
\usepackage{listings}%

\theoremstyle{thmstyleone}%
\theoremstyle{thmstyletwo}%

\theoremstyle{thmstylethree}%

\hypersetup{
pdftitle={OpenHealth Lake: Designing and testing a data lakehouse platform for health applications},
pdfauthor={Marcel Dunaiski},
pdfkeywords={Jailbreaking, Large Language Models, Low-resource languages, LLMs},
}

\begin{document}

\title{OpenHealth Lake: Designing and testing a data lakehouse platform for health applications}


\author{ \href{https://orcid.org/0000-0001-5740-3968}{\includegraphics[scale=0.06]{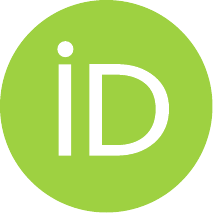}\hspace{1mm}Danilo Silva} \\
	Computer Science Division \\
	Mathematical Sciences Department\\
	Stellenbosch University\\
	Stellenbosch, South Africa \\
	\texttt{27800164@sun.ac.za} \\
	\And
	\href{https://orcid.org/0000-0003-1095-1910}{\includegraphics[scale=0.06]{orcid.pdf}\hspace{1mm}Monika Moir} \\
	Center for Epidemic Response and Innovation \\
	Stellenbosch University \\
	Stellenbosch, South Africa \\
	\texttt{monikamoir@gmail.com} \\
	\And
	\href{https://orcid.org/0000-0002-6033-7655}{\includegraphics[scale=0.06]{orcid.pdf}\hspace{1mm}Cheryl Baxter} \\
	Center for Epidemic Response and Innovation \\
	Center for the AIDS Programme of Research in South Africa (CAPRISA) \\
	\And
	the Inform Africa Research Study Group \\
	\texttt{contact@inform-africa.org} \\
	\And
	\href{https://orcid.org/0000-0002-3027-5254}{\includegraphics[scale=0.06]{orcid.pdf}\hspace{1mm}Tulio de Oliveira} \\
	Center for Epidemic Response and Innovation, Stellenbosch University and \\
	Center for the AIDS Programme of Research in South Africa (CAPRISA) and \\
	KwaZulu-Natal Research Innovation and Sequencing Platform (KRISP) and \\
	Department of Global Health, University of Washington, United Stated\\
	\And
	\href{https://orcid.org/0000-0002-4649-6270}{\includegraphics[scale=0.06]{orcid.pdf}\hspace{1mm}Joicymara Xavier} \\
	Center for Epidemic Response and Innovation and\\
	Computer Science Division, Department of Software and System Information \\
	Instituto Tecnológico de Aeronáutica (ITA), Brazil \\
	\texttt{joicymara@ita.br} \\
	\And
	\href{https://orcid.org/0000-0003-1957-3979}{\includegraphics[scale=0.06]{orcid.pdf}\hspace{1mm}Marcel Dunaiski} \\
	Computer Science Division, Mathematical Sciences Department\\
	Stellenbosch University, Stellenbosch, South Africa \\
	\texttt{marceldunaiski@sun.ac.za} \\
}

\maketitle

\begin{abstract}
Data management can be a complex challenge in fields such as bioinformatics and health sciences, which continuously generate extensive heterogeneous datasets. In the context of collaborative global health initiatives, secure storage and sharing of data are crucial to support impactful research. However, the absence of a unified data management platform complicates efficient data exchange and governance within these initiatives. In this paper, we introduce the design process of OpenHealth Lake, a data management prototype platform based on a data lakehouse architecture, data federation, and the FAIR principles. The platform is designed using open-source tools, guided by system requirements identified in previously published studies and complemented by insights from the existing literature. The current prototype platform comprises a user-friendly website, an open API, Python and R packages, allowing users to interact with the platform in multiple ways. Through a user study that included participants with varying technical backgrounds, we showed that our proposed data management prototype is both usable and useful. Our prototype design showcases the adaptability, scalability, and reproducibility of a lakehouse system that can be used by any organisation. It is designed as a flexible and complementary approach that allows organisations to customise data management systems to their specific requirements and resources, including cloud-based or self-hosted storage choices.
\end{abstract}

\keywords{Data management, Governance Software, Lakehouse, Data Lake, Open-source, Databases}




\section{Introduction}

The rapid digital innovation and growth of new technologies, such as Next Generation Sequencing, have resulted in the generation of extensive heterogeneous data volumes~\citep{o2013big}. Data has been characterized as a key resource for new advancements and breakthroughs not only in genomic research, but in science and business-related areas in general \citep{schneider2024lakehouse}. Simplified data management solutions and paradigms that focus on structured data were the convention until recent years. However, due to the exponential data generation, they are no longer sufficient to provide flexible, scalable, and cost-effective data management solutions~\citep{o2013big, schneider2023assessing}.

Biological research often features collaboration between multiple organizations around the world. As a result, the output data is highly decentralized and must comply with various data regulations. The absence of solutions that offer distributed storage interconnected with centralized access poses a significant challenge for research collaboration \citep{zhang2011biomart}. The emergence and adoption of databases featuring data federation capabilities have facilitated extensive and responsible data sharing by enforcing data governance standards for access control, provenance, and privacy. In a federated approach, multiple data sources are connected and integrated through a relational database system that serves as middleware, providing unified and transparent access to the different sources while maintaining full control over user access~\citep{haas2002data}. However, federated databases built on traditional structured data management capabilities may not be sufficient to scale with large, heterogeneous, and unstructured data generated in fields such as bioinformatics.

The data lakehouse paradigm has emerged as a distinct solution for managing heterogeneous data, which combines the capacity to handle large volumes of unstructured data with structured data storage and analytical capabilities~\citep{armbrust2021lakehouse, harby2022From, schneider2023data, silva2025review}. It benefits from diversified storage systems (such as file systems, object storage, and non-relational databases) and processing technologies (such as Apache Spark and Delta Lake) to enforce scalability and flexibility, while eliminating the need for complex data Extraction, Transformation, and Loading (ETL) pipelines to handle incoming data. Nevertheless, lakehouse solutions remain unexplored in science-related domains, as many existing implementations primarily depend on traditional paradigms. Therefore, the combination of the lakehouse principles with the data federation approach holds significant potential to address many of the existing challenges~\citep{silva2025review}.

The Role of Data streams In Informing infection dynamics in Africa (INFORM Africa) research hub was created to focus on developing strategies for effectively managing health-related research data from Nigeria and South Africa~\citep{poongavanan2023managing}. INFORM’s main challenges lie not only in storing and processing large volumes of data but also in establishing effective data governance and access control policies. To address the challenges faced in the hub while overcoming the shortcomings of traditional data management solutions, we previously envisioned the data lakehouse as an alternative approach to meet its data management needs~\citep{silva2025review}.

Assessing software usability is a crucial step in the development of solutions aimed at addressing data management gaps, as user-friendly features often determine whether a new system will achieve widespread adoption. Software quality can be evaluated based on various criteria that are specific to the application domain of the software. In general, some criteria can be used to formulate an assessment framework, including functionality, reliability, usability, efficiency, maintainability, and portability~\citep{sagar2017systematic}. Efficiency, reliability, and effectiveness are key criteria for the success of a data lakehouse solution, but designing assessment studies to accurately evaluate solutions poses a significant challenge due to the absence of standards to measure these criteria~\cite{sawadogo2021benchmarking}.

Here, we present the design process of OpenHealth Lake, a data lakehouse prototype platform to address common challenges in managing heterogeneous data. Based on the findings of our review study~\citep{silva2025review}, for which the INFORM Africa research hub served as the use case, we collected insights that guided this prototype's design.  In addition, we conducted a user study to evaluate the effectiveness and usability of the prototype's core functionalities.


\section{Prototype Implementation}

The design process of a data lakehouse solution for data management can be challenging and often unclear, as this approach tries to merge the best features of both data lakes and data warehouses, increasing complexity in the development process~\citep{janssen2024evolution}. The implementation of OpenHealth Lake, our data lakehouse prototype platform, was based on the architectural concepts found in the literature, along with our own review of current tools and strategies to design it~\citep{silva2025review, harby2022From, armbrust2021lakehouse, terrizzano2015data}. 

Developing a federated approach with data flexibility and scalability is vital to address the majority of challenges in heterogeneous data management systems. Tools that provide distributed and schemaless capabilities, such as object storage and distributed non-relational databases, have the potential to be integrated in a broader architecture to bridge the data management gaps identified in our previous study~\citep{silva2025review}. The integration between such tools with custom application interfaces and cataloguing strategies in a robust architecture has the potential to effectively provide data storage, governance, and analytics, while addressing the system requirements for data management architectures.

\section{Data lakehouse prototype architecture}

\begin{figure*}[!t]%
\centering
\includegraphics[width=0.8\linewidth]{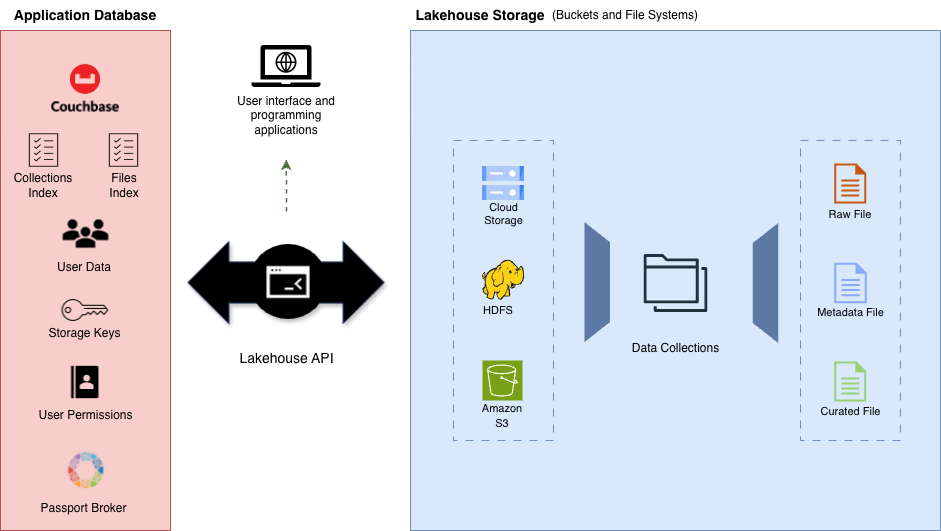} 

\caption{Lakehouse Infrastructure design. The system comprises three main modules: the application database, the programming interface (API), and the file storage layer. Couchbase, a non-relational database, stores metadata and application management data such as user credentials, storage keys, data collection and file indexes, and user permissions. On the storage side, systems like Google Cloud Storage, Amazon S3, and HDFS manage both structured and unstructured data files. The Lakehouse API bridges the application database and storage systems, providing programmatic and user interface access to the entire platform.}

\label{fig:lakehouse_structure}
\end{figure*}

To align OpenHealth Lake with data federation, FAIR principles, and the common data management system requirements, we structured its architecture into three layers: data ingestion, data storage, and data governance~\citep{barker2022introducing, silva2025review}. The ingestion and governance layers are implemented via a custom Representational State Transfer (REST) API using Python (v$3.12$) and FastAPI (v$0.111$), with endpoints to manage user accounts, data uploads and downloads, as well as data catalogue operations to handle metadata indexing from files and collection of files. Furthermore, the API is designed to expose all functionalities from the lakehouse platform to authenticated external apps, allowing for external interaction with the data in the storage layer and data management methods. Figure~\ref{fig:lakehouse_structure} illustrates the infrastructure design.

The lakehouse storage layer was designed to be both data-agnostic and flexible. Raw and processed data can be seamlessly stored in this layer without requiring preprocessing steps or ETL pipelines. This layer was implemented using a hybrid approach that enables access to cloud buckets and self-hosted Hadoop Distributed File System (HDFS) clusters, to fulfil the flexibility requirements to accommodate different data governance conditions.

In our proposed lakehouse architecture, one or more files can be grouped in a logical file repository defined as a file {\em collection}, which works analogously to a git file repository that groups and aggregates code and data files into a logical repository/project that can be managed to restrict user access and to create other top-level governance strategies over the group of files in it. When a collection is created, its application metadata is stored in a document record in the database. We refer to this group of documents in the database as the {\em collection index}. Similarly, when a new file is uploaded into the platform, it must be linked to a given collection (e.g., file repository). Uploaded files have their metadata extracted and stored in the database to serve as an index of files. The collection index together with the file index form the {\em data catalogue}, which is logically designed to support data discovery in OpenHealth Lake (see Figure~\ref{fig:lakehouse_structure}).

Instead of MongoDB, which was suggested in our previous paper~\citep{silva2025review}, we selected Couchbase to manage data related to the OpenHealth Lake's application routines (e.g., user credentials, storage access keys, access requests) and dataset indexes. Similar to MongoDB, Couchbase (developed by Couchbase Inc.\footnote{\url{https://www.couchbase.com/}}) is also a non-relational, schema-agnostic, open-source, and distributed database system. However, Couchbase offers some advantages over MongoDB. Firstly, it combines features from key-value and document-oriented non-relational paradigms, rather than being purely document-oriented~\citep{couchbase_whynosql, couchbase_why}. Secondly, its built-in memory-first caching architecture allows for cache management at the database bucket level, which enables Couchbase to serve as an application-level caching layer on its own~\citep{couchbase_why}. In contrast, MongoDB relies on operating system-level caching, which improves performance but is not designed primarily as an application-level (database-level) cache~\citep{mongodb_wiredtiger_manual, mongodb_storage_faq}. Couchbase also provides an intuitive user interface that facilitates the setup of distributed clusters with multidimensional scaling, which enables independent memory quota allocation for each functionality of the database (e.g., storage, indexing, querying, analytics, full text search, and eventing).

\begin{figure*}[!t]%
\centering
\includegraphics[width=0.85\linewidth]{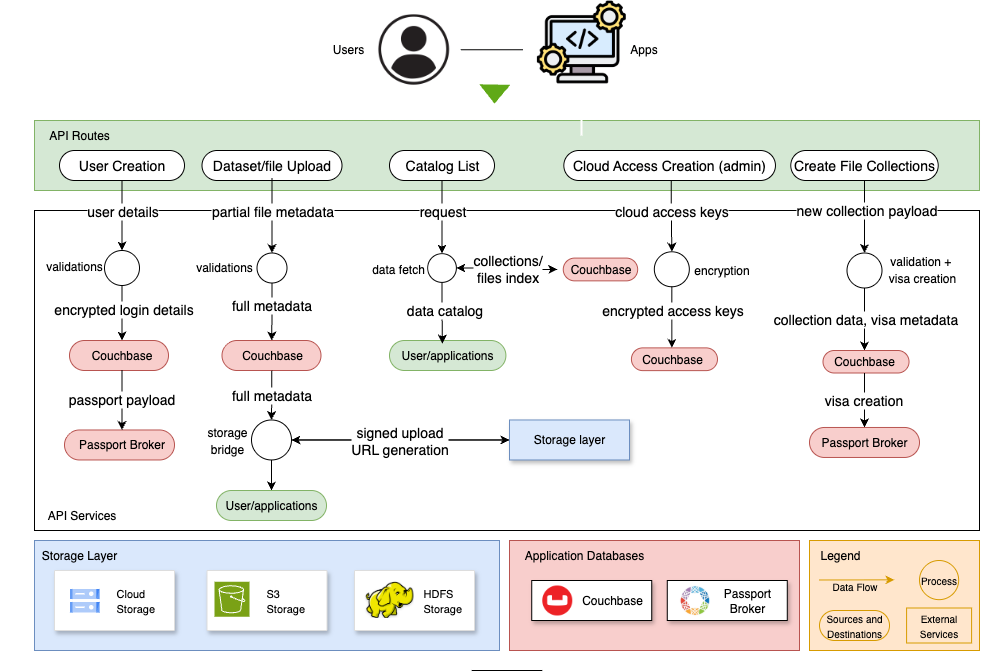} 

\caption[Simplified API data flow diagram.]{Simplified API data flow diagram. The diagram illustrates the operations and data flows of each core API endpoint. Each endpoint includes a data validation step prior to processing and storage. User login credentials are stored in Couchbase, while data passports (e.g., access permissions) are stored in both Couchbase and Passport Broker. The Storage Layer is used to store structured and unstructured files.}

\label{fig:use_cases_diagram}
\end{figure*}

The data stored in Couchbase is controllable through OpenHealth Lake's REST API, with which users can insert and retrieve data from the database. Login credentials are stored in a record in the Couchbase database (see Figure~\ref{fig:lakehouse_structure}) when a new user is created, while their dataset access permissions and requests are recorded whenever the access request functionality is invoked. File metadata is stored in the file's index during file uploading operations. Cloud bucket and self-hosted HDFS access keys are stored in a Couchbase document with encrypted values to keep users' access keys protected even if the database is exposed. Figure~\ref{fig:use_cases_diagram} shows a simplified data flow diagram of the API's core functionalities. The platform's database models can be visualized in the Figure~\ref{fig:database_model}, in the appendix.

\section{API design}

We used Python's FastAPI framework to build OpenHealth Lake's REST API, which is a simple yet robust solution for high-performance API development. Furthermore, we used Docker to simplify the API deployment process by providing an isolated container environment, reducing the need to manually configure the host machine each time the platform is deployed on a new server. This approach further ensures that the platform's deployment aligns with the FAIR4RS principles, particularly those concerning reusability and interoperability. 

\begin{figure*}[ht]
    \centering
    \includegraphics[width=0.8\linewidth]{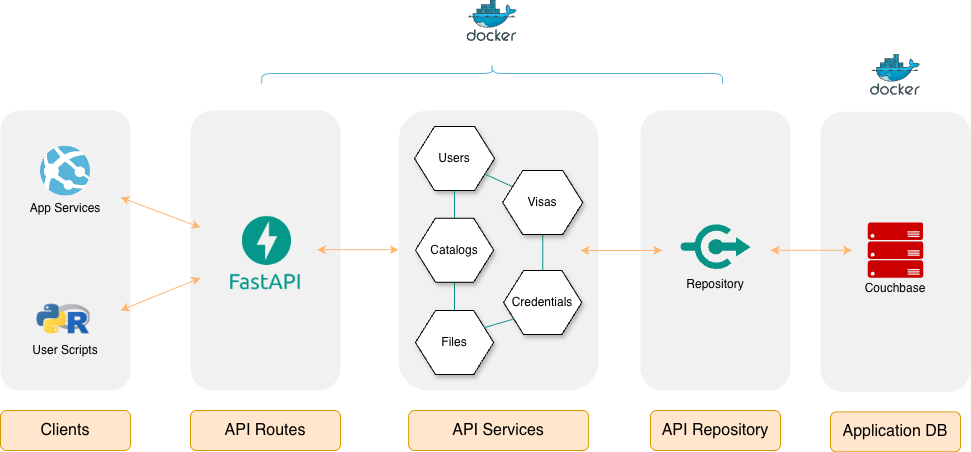}
    \caption[API infrastructure diagram.]{API infrastructure diagram. This diagram shows the OpenHealth Lake's API design and its main services. Each service is associated with a key functionality of a generic data management application, identified from the literature and in our previous study.}
    \label{fig:api-infra-diagram}
\end{figure*}

OpenHealth Lake's REST API is structured into three layers: the API routes layer, the API services layer, and the API repository layer. The routes layer implements the FastAPI endpoints that handle interactions with the API. The services layer implements the functionalities derived from generic user requirements, while the repository layer establishes the connection with the application database and exposes basic operations such as data insertion, retrieval, update, and deletion. The Builder pattern was applied to implement helper classes for database query construction, whereas the Facade pattern was used to manage interactions between the API routes (endpoints) and other components, such as the services and repository layers. Figure~\ref{fig:api-infra-diagram} illustrates the REST API application structure. 

The \textbf{User Services} of the API services layer expose all functionalities related to managing users within OpenHealth Lake, such as user creation, user updates, user deletion, and user password recovery. These services encrypt passwords when creating a new user and store them along with other user details in the application database. The user details and passwords are utilised for authentication, login, and data access control based on user access permissions. The API application makes a clear distinction between user login credentials and user access permissions. Login credentials refer to a user's login and supplementary information, such as email, password, role, and ID, whereas access permissions refer to the mechanism governing dataset access within the entire platform.

We implemented a passport strategy for data access control that integrates the GA4GH passport broker service. With this strategy, users interact with datasets by means of granted visas. When a new dataset collection is created on the platform, a corresponding visa is automatically generated by the \textbf{Visas Service}. Visas can then be granted to, or revoked from, individual users by the collection's owners (publishers). The Visa Services are responsible for controlling visa creation, update, deletion, granting, and revoking functionalities, as well as storing the users' visas in the passport broker.

The \textbf{Catalog Service} provides comprehensive methods for managing collections of files and file records in Couchbase, enabling data discovery and indexing. Users may browse the collection index to find and request access to private collections of files. They may also use the file index to find different versions of files that can be downloaded on the platform.

The \textbf{File Service} manages all file upload and download operations within OpenHealth Lake. To support data provenance and prevent duplication, the method \texttt{upload\_file} incorporates an automated version control mechanism aligned with the system requirements. When an incoming file matches an existing file based on stored metadata (e.g., file name, file collection, file category, and bucket location), the service assigns a new version prefix rather than overwriting or rejecting the upload. This prefix reflects the sequential version of the file (e.g., \texttt{collection\_name/v$2$/file\_name}, \texttt{collection\_name/v$3$/file\_name}, or \texttt{collection\_name/v$n$/file\_name}), ensuring that each revision is traceable and uniquely identifiable. 

OpenHealth Lake also supports manual version assignment during file upload, allowing users to record externally processed or independently versioned files explicitly. This feature is especially valuable in collaborative and federated research environments, where datasets may undergo preprocessing, annotation, or revision before ingestion. By enabling manual control over version labels, the platform ensures that all revisions remain traceable and correctly associated with their provenance. All file metadata records with file version can be retrieved from the application database.

The API currently supports connections with three different file and storage systems, namely, HDFS, Amazon S3, and Google Cloud Storage. Thus, the File Service is not only responsible for handling file uploads and downloads, but also for selecting the correct storage environment by generating direct upload URLs for the required storage system. Different HDFS clusters or cloud storage buckets can be included in the storage zone through the \textbf{Credential Services}. These services manage the access keys that grant access to external storage environments, allowing data managers to connect the platform to an unlimited number of external data sources by registering new access keys, which are securely encrypted and stored in Couchbase. The encryption step was included as an additional security mechanism in the event of database breaches, ensuring that not only user passwords but also external data source keys controlled by independent entities remain protected. This step is implemented using a symmetric encryption routine named \texttt{Fernet}\footnote{\url{https://cryptography.io/en/latest/fernet/}}, from the library \texttt{cryptography}, where a single secret key is used to both encrypt and decrypt the data. The secret key must be defined by the data manager when deploying OpenHealth Lake.

To further enhance accessibility and interoperability, we developed libraries for Python\footnote{\url{https://github.com/danilo-dcs/lakehouse-python-package}} and R\footnote{\url{https://github.com/danilo-dcs/lakehouse-R-package}} environments. These libraries encapsulate the API's core functionalities and simplify their use, enabling users with basic programming knowledge and limited software engineering skills to directly integrate the lakehouse into their analysis scripts. These libraries were used in our usability study to measure the effectiveness and usability of our prototype.

\section{Usability Study Methods}

Assessing software usability is a crucial step in the development of solutions aimed at addressing data management gaps, as user-friendly features often determine whether a new software will achieve widespread adoption. Software quality can be evaluated based on various criteria that are specific to the application domain of the software. In general, some criteria can be used to formulate an assessment framework, including functionality, reliability, usability, efficiency, maintainability, and portability~\citep{sagar2017systematic}.

We designed a user study to evaluate the main functionalities offered by OpenHealth Lake through its Python and R libraries. We also assessed users' perception and the time required to become familiar with the libraries. For the user study, we recruited participants with different educational levels from various fields, including bioinformatics, microbiology, biology, computer science, and data science. The recruitment process and usability study protocol received ethical approval from the REC: SBE ethics committee at Stellenbosch University (South Africa) under project ID 30945.

The designed study comprised four main components:
\begin{enumerate}
\item A standardised introductory presentation about our implemented prototype,
\item an informed consent form, 
\item the main user study with eleven tasks (questionnaires) to be completed using our system, and 
\item a final feedback form.
\end{enumerate}

The introductory presentation introduced the overall structure of our data management platform and its purpose, highlighting the key features and core functionalities. Moreover, we provided a brief introduction about the process of the usability test, which contained instructions for the informed consent form, the main task-solving session, and the final feedback form. Components $2$, $3$, and $4$ were implemented using Microsoft Forms with preconfigured anonymous forms to collect participant responses. The forms were designed to guide the participant through an ordered sequence of task questionnaires, requiring each questionnaire to be submitted before accessing the subsequent task. The participants were not permitted to move back to the completed questionnaires. Each questionnaire included a ``skip'' option to allow participants to proceed in cases where they encountered difficulties completing a task.

The tasks were constructed to evaluate user interactions with all functionalities provided by our prototype platform using either Python or R (see Table~\ref{tab:tasks} for an overview of the tasks). Each task was associated with a distinct functionality and was accompanied by a set of questions about the data and information stored in the prototype's database. Participants were instructed to use our prototype's documentation to identify and execute the most appropriate functions needed to complete each task and answer the questions. The questionnaires included descriptive, task-oriented questions such as: {\em ``Search for any files that contain the keyword \texttt{sequences} in their file name. How many files appear in the result list?''} or {\em ``Fetch the data collections list, and answer: how many collections are there?''}.

Lastly, the final feedback form collected background information about the participants, including their years of experience, education level, and field of work. Furthermore, in this form, participants rated the usability and usefulness of the tested functionality associated with each task using a Likert scale with the following options: Very difficult, Difficult, Neutral, Easy, Very Easy, Useful, Not Useful.

\begin{table}[h]
\begin{threeparttable}
\caption[Overview of tasks of the user study.]{The list of tasks of the user study's questionnaire component with the associated prototype functionalities tested.}
\label{tab:tasks}
\begin{tabular*}{\columnwidth}{@{\extracolsep\fill}lll@{\extracolsep\fill}}
\toprule
Task & Tested functionality  \\
\midrule
Task $0$ & Setup/Install \\
Task $1$ & Authentication  \\
Task $2$ & Listing collections   \\
Task $3$ & Listing files    \\
Task $4$ & Searching files \\
Task $5$ & Fetching dataframe\\
Task $6$ & Downloading file\\
Task $7$ & Listing buckets\\
Task $8$ & Creating collection\\
Task $9$ & Uploading file\\
Task $10$ & Dataframe upload\\
Task $11$ & Data access request\\
\bottomrule
\end{tabular*}
\begin{tablenotes}%
\item Note: Tasks $0$, $1$, and $7$ had auxiliary purposes and did not test critical system functionalities, but rather guided the user towards the next task.
\end{tablenotes}
\end{threeparttable}
\end{table}

After the usability study was conducted, we processed the collected data and aggregated it to obtain results on user responses and metrics such as time spent per task and dropout rates. The data cleaning step included manual harmonisation of the respondents' text answers, calculation of the completion times, and dropout rates per task.




\section{Usability Study Results and Discussion}



    

Our usability study was designed to be simple yet informative. The responses collected provided valuable information on the effectiveness of the prototype by recording task completion times, dropout rates, and overall user satisfaction. The study involved $31$ participants, of whom $26$ completed all tasks. Our participants' attribute analysis revealed that $11$ participants hold a Postgraduate degree, $12$ have a PhD degree, and only one participant has an Undergraduate degree (Table~\ref{tab:profile}). Most participants had more than two years of work experience, indicating limited involvement from early-career individuals in our study.


\begin{table}[!h]
\caption{Total number of participants per experience group and educational level.}
\label{tab:profile}

\begin{tabular*}{\columnwidth}
{@{\extracolsep{\fill}}rrrr@{\extracolsep{\fill}}}
\toprule
Experience \\ group (years) & Undergraduate & Postgraduate & PhD \\
\midrule
$0-2$    & - & 1 & 2 \\
$3-5$    & - & 6 & 3 \\
$6-8$    & - & 1 & 1 \\
$9-11$   & - & 2 & 3 \\
$12-14$  & 2 & 2 & 1 \\
$18-20$  & - & - & 1 \\
\bottomrule
\end{tabular*}

\begin{tablenotes}
\item \footnotesize Note: Postgraduate includes Master's and MBA. A dash (-) indicates no participants in that category.
\end{tablenotes}
\end{table}

\begin{figure}[h]%
\centering
\includegraphics[width=\linewidth]{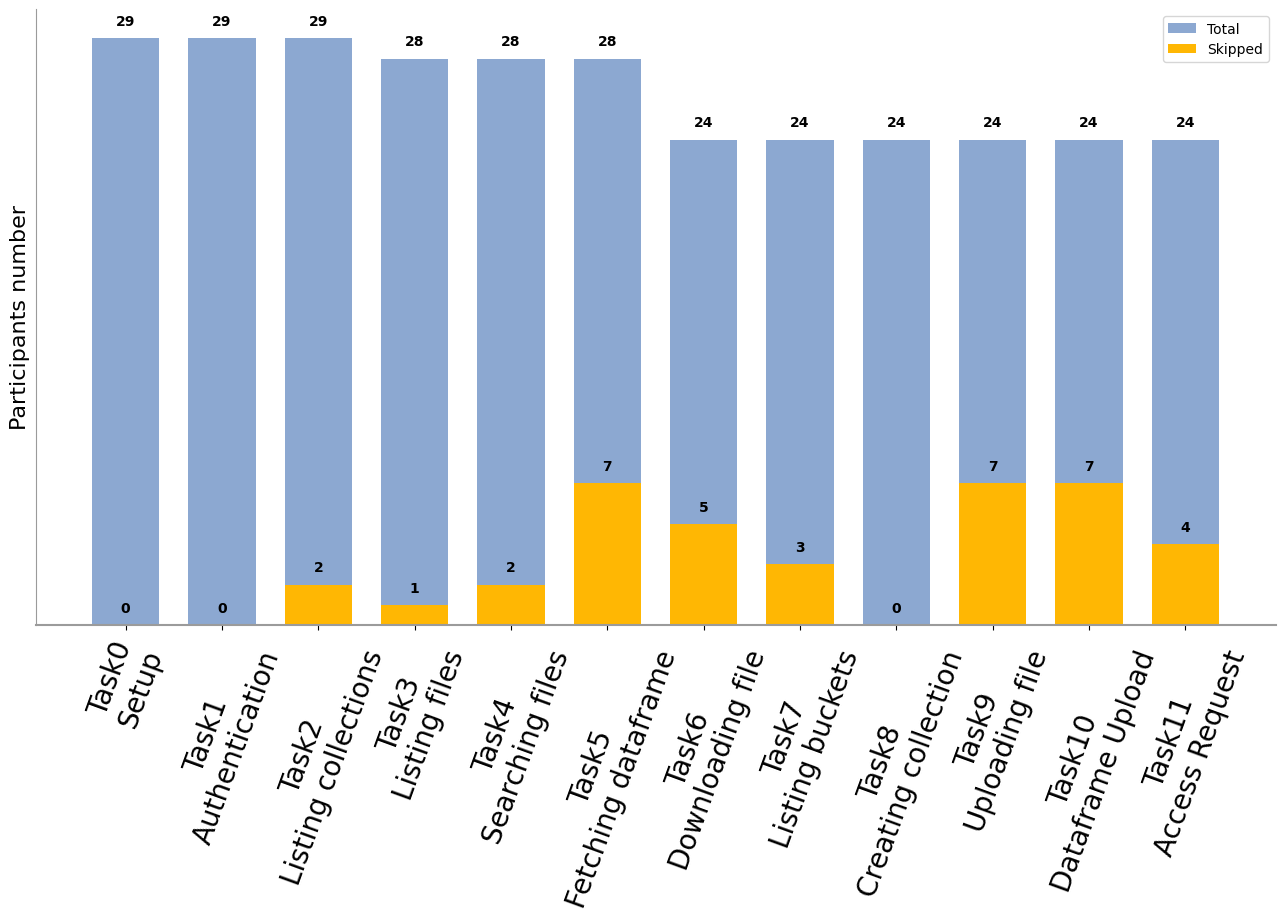} 

\caption{Total number of participants versus number of participants who skipped each task. No participants skipped the initial tasks, but the number of incomplete tasks began to increase from task two, reaching peaks at tasks five, nine, and ten.}

\label{fig:skip_number}
\end{figure}

The participants exhibited a high level of engagement during the task-solving phase of the study, with only five participants withdrawing before completing all tasks, resulting in a dropout rate of $17\%$ (see the Figure~\ref{fig:dropout_rate} in the appendix with the dropout rate curve through the study tasks). However, analysis of participants’ behaviour per task revealed that many were unable to complete certain tasks and were compelled to skip them. Tasks~5, ~6, ~9, and 10 had the highest rate of incompletion, as illustrated in Figure~\ref{fig:skip_number} (see Table~\ref{tab:tasks} for the task list).

Fourteen participants chose to perform the usability study with our Python library, and fifteen chose R, which provided a balanced split for the analysis. In total, we conducted four user study sessions with distinct participants. R was the predominant library chosen in the first session, which mainly comprised biologists, bioinformaticians, and participants from other health-related fields. In the subsequent sessions, Python was the predominant choice among participants, who were primarily computer scientists and data scientists. This suggests that users with health or biological backgrounds tend to be more comfortable with R, whereas those from computer science are likely more familiar with Python. This distinction is likely due to their exposure to these environments during their professional or academic careers. This observation confirms our decision to support tools in both programming languages, since it increases accessibility and usability across diverse user backgrounds.


Our study revealed that the users who chose to use Python found our prototype implementation generally more usable compared to the R users. Our results showed that $47\%$ of the users who rated the tasks, and the usability of their related prototype's functionalities, as ``Very Easy'' or ``Easy'', were using Python, whereas $41\%$ were using R. Furthermore,  the number of participants who used Python and ranked the overall tasks as ``Neutral'' or ``Difficult'' accounted for less than one percent, while $11\%$ of the participants that used R and rated the tasks as ``Neutral'' or ``Difficult''. Table~\ref{tab:responses} shows the complete results from the usability questionnaire. The reasons for the performance differences may be: (1) our Python library was better designed and offers better usability, (2) the Python language environment is easier to use for users with limited coding experience given its relatively gentle learning curve, (3) the participants' background played a key role in their performance, or (4) both Python and R libraries' usability paradigm influenced the participants performance. 

\begin{table}[h!]
\caption[Percentage of usability responses per coding environment.]{Percentage of usability responses per coding environment. These responses were collected through the final feedback form, which asked users to explicitly rate each task and its associated prototype functionality as ``Very Easy'', ``Easy'', ``Neutral'', ``Difficult'', or ``Very Difficult''. The percentages shown represent the aggregated results of all responses grouped by usability category.}
\label{tab:responses}
\begin{tabular*}{\columnwidth}{@{\extracolsep\fill}llll@{\extracolsep\fill}}
\toprule
Response & Python & R  \\
\midrule
Very Easy & $33.60$\% & $30.83$\% \\
Easy	& $13.44$\% & $10.67$\% \\
Neutral & $0.79$\% & $8.70$\%  \\
Difficult & - & $1.98$\%  \\
Very Difficult & - & - \\
\bottomrule
\end{tabular*}
\end{table}

We argue that the first reason is only a primary driver of the observed performance differences if the libraries had markedly different structures. However, both provide equivalent functionalities, structures, and documentation designs. The second reason appears less plausible, as most participants possessed at least basic coding expertise. Therefore, language learning was unlikely to be the main factor influencing performance. The third reason seems to be more likely, given that most Python users were computer scientists with greater experience in Python and programming principles in general. The fourth reason, in combination with the third, may also have contributed, as the libraries were designed as wrappers for OpenHealth Lake's REST API interfaces, following a web service-oriented approach. Participants in the R-user group may have been proficient in data processing, curation, and cleaning, but were less experienced with using libraries that follow a web-service usability paradigm, such as instantiating a client object and using CRUD-like library methods.

\begin{figure}[h]%
\centering
\includegraphics[width=1\linewidth]{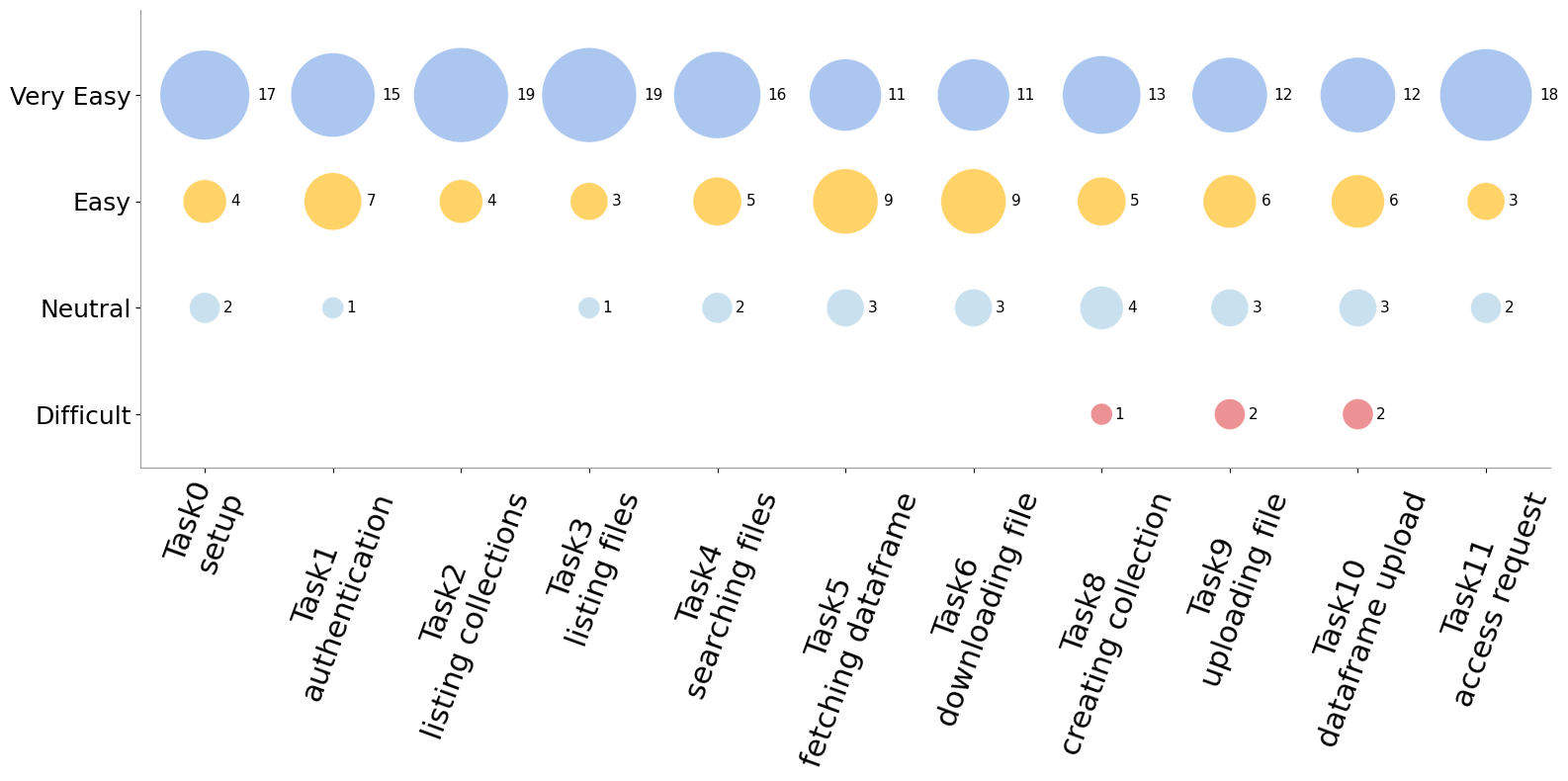} 

\caption[Usability Responses based on user perception per task.]{Usability responses based on user perception per task. The bubble chart illustrates the total number of responses across four usability categories: ``Very Easy'', ``Easy'', ``Neutral'', and ``Difficult'', in accordance with the final feedback form. Most responses rated the tasks as ``Very Easy'' or ``Easy'', while a small portion of participants rated certain tasks as ``Neutral'' or ``Difficult''.}

\label{fig:usability_per_task}
\end{figure}

To measure user performance in each task, we analysed the usability responses from the final feedback form and compared them with the task completion times. The results showed that, regardless of the chosen programming language, participants rated most tasks as ``Easy'' or ``Very Easy'', accounting for $89\%$ of the responses. ``Neutral'' accounted for $10\%$, while ``Difficult'' represented only $2\%$. The distribution of usability responses per task, given in Figure~\ref{fig:usability_per_task}, highlights that the tasks related to the functionalities of creating collections, uploading files, and uploading dataframes were the only ones perceived as ``Difficult''. This suggests that these functionalities may be the least intuitive. All tasks also received ``Neutral'' responses, indicating that some users either skipped them or completed them without perceiving them as particularly ``Easy'' or ``Difficult''.

Tasks~8, 9, and~10 are linked to functionalities that require more input arguments to be specified than most of the previous functionalities (see Table~\ref{tab:tasks} for the task list). The \texttt{collection\_name}, \texttt{storage\_type} (storage environment), and \texttt{bucket\_name} (physical location for the collection) are required parameters to create a new collection, while the \texttt{local\_file\_path}, \texttt{final\_file\_name}, \texttt{file\_category}, and \texttt{collection\_catalogue\_id} are required for uploading a new file. 

In terms of the parameters, the \texttt{storage\_type} and the \texttt{bucket\_name} required in Task 8 are provided in the output of Task 7, forcing the participants to look back at the previous tasks' output to complete it. The \texttt{local\_file\_path} argument must specify the location of the target file on the host machine. During the usability sessions, we observed that some R users using Windows operating systems encountered issues, since our R library did not convert paths accordingly. For example, a declared path \texttt{files/sample/sequences.fasta} should be converted to \texttt{files\textbackslash sample\textbackslash sequences.fasta)}. Users were also confused about the \texttt{final\_file\_name} parameter, as they expected files to be stored in OpenHealth Lake under their original names rather than declaring a new name manually. The \texttt{collection\_catalogue\_id} was straightforward to understand, as it indicates the target data collection in the platform where the file must be stored.

\begin{figure}[h]%
\centering
\includegraphics[width=1\linewidth]{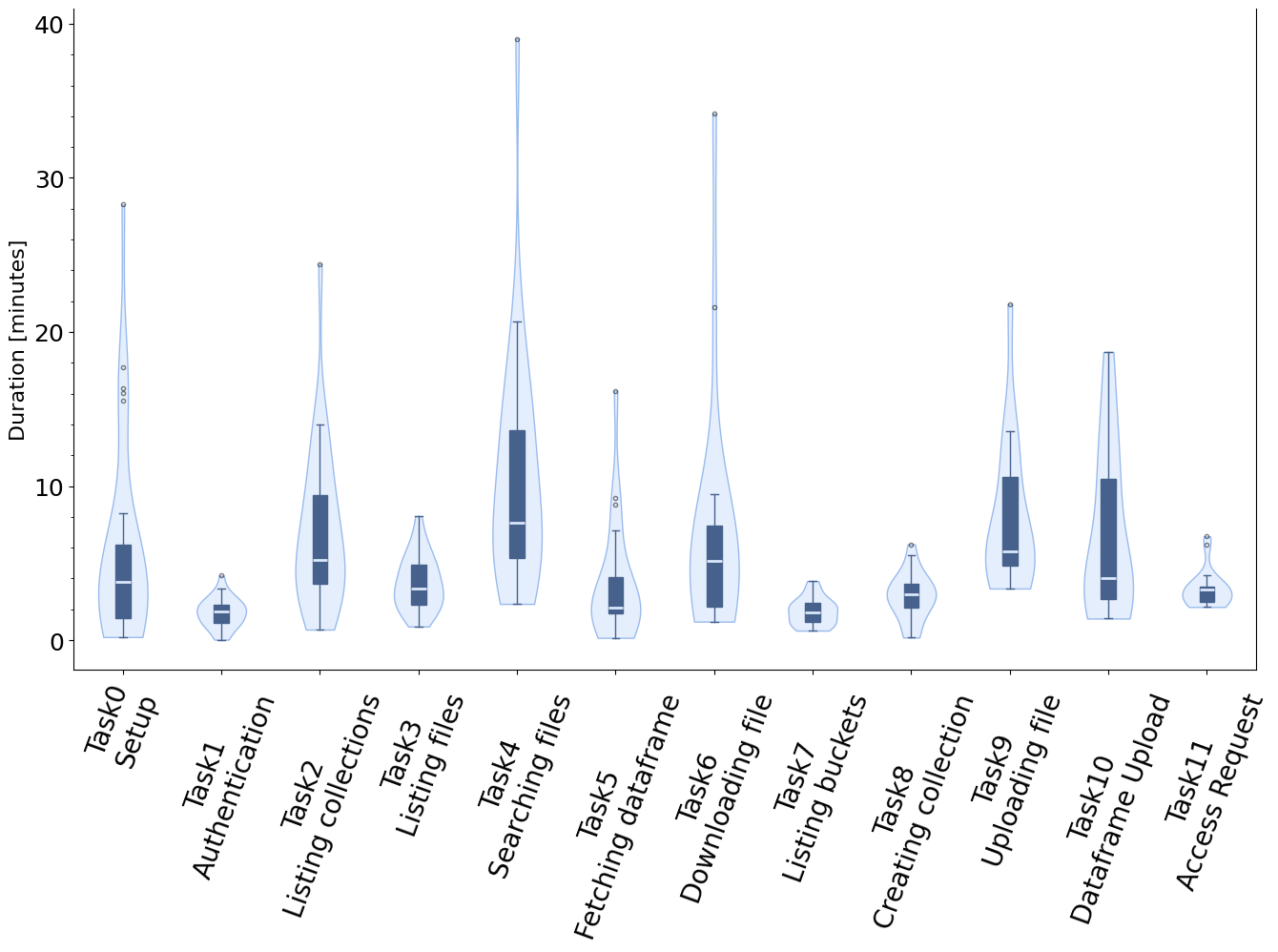} 

\caption[Distribution of time spent per task.]{Distribution of time spent per task. Each violin shape represents the spread and density of participants' task durations, while the box plot inside indicates the median and interquartile range.}

\label{fig:time_taken_per_task}
\end{figure}

Although users perceived the tasks as ``Easy'' or ``Very Easy'', the completion time results presented in Figure~\ref{fig:time_taken_per_task} suggest that participants required a considerable amount of time to orient themselves with the documentation and to complete six of the 12 tasks. Conversely, the distribution of time spent to complete Tasks 8, 9, and 10 was not the highest, despite some participants classifying them as ``Difficult'' (see Table~\ref{tab:tasks} for the task list). Participants spent the longest time completing task 4, likely due to its functionality characteristics.

Tasks 0 and 1 were designed purely for guidance purposes, providing participants with detailed instructions and complete code to install the library and perform authentication. In contrast, the remaining tasks included descriptions and questions, without hints for their solution, thus requiring participants to consult the documentation. The outliers observed in Figure~\ref{fig:time_taken_per_task} can be attributed to users who spent additional time interpreting the task descriptions or interacting with the documentation, which may have diverted their focus from the task questionnaire. Additional comments were provided by the participants in the final feedback form, highlighting difficulty in understanding the task flow in the beginning, the documentation format, and the included code examples.

Task 4 was divided into two components: Basic File Search and Advanced File Search. In the Basic File Search component, participants were required to use a library function to retrieve files based on a keyword input. This function returns a list of files whose names partially correspond to the specified keyword. The second component involved executing an advanced search via our library's query function, which requires as input a set of strings containing search parameters (for example, \texttt{file\_name=zika.csv}, \texttt{file\_category=structured}). These functions were found to be less intuitive and straightforward for users with limited experience in database querying, thereby necessitating closer consultation of the documentation and adaptation of the provided example code.

Task 8 was also designed for guidance, as it does not evaluate any core functionality, but rather an auxiliary functionality to guide the participants in obtaining the information required to solve the following task (Task 9).

The median and maximum times spent on Task 2 (listing collections) were higher than expected. Participants were required to access the documentation and locate the relevant code example to execute and complete the task. Nevertheless, this functionality does not require any additional parameters, which means that the example code provided in the documentation is sufficient to obtain the correct output and answer the questions. Lastly, the time taken to complete Task 11 was among the lowest, suggesting high usability and ease of use. It is important to note that Task 11 was the only task performed through the website, requiring no coding skills. This task specifically evaluated the data access request functionality, which is exclusively available on the website associated with our prototype implementation.

\begin{figure}[h]%
\centering
\includegraphics[width=1\linewidth]{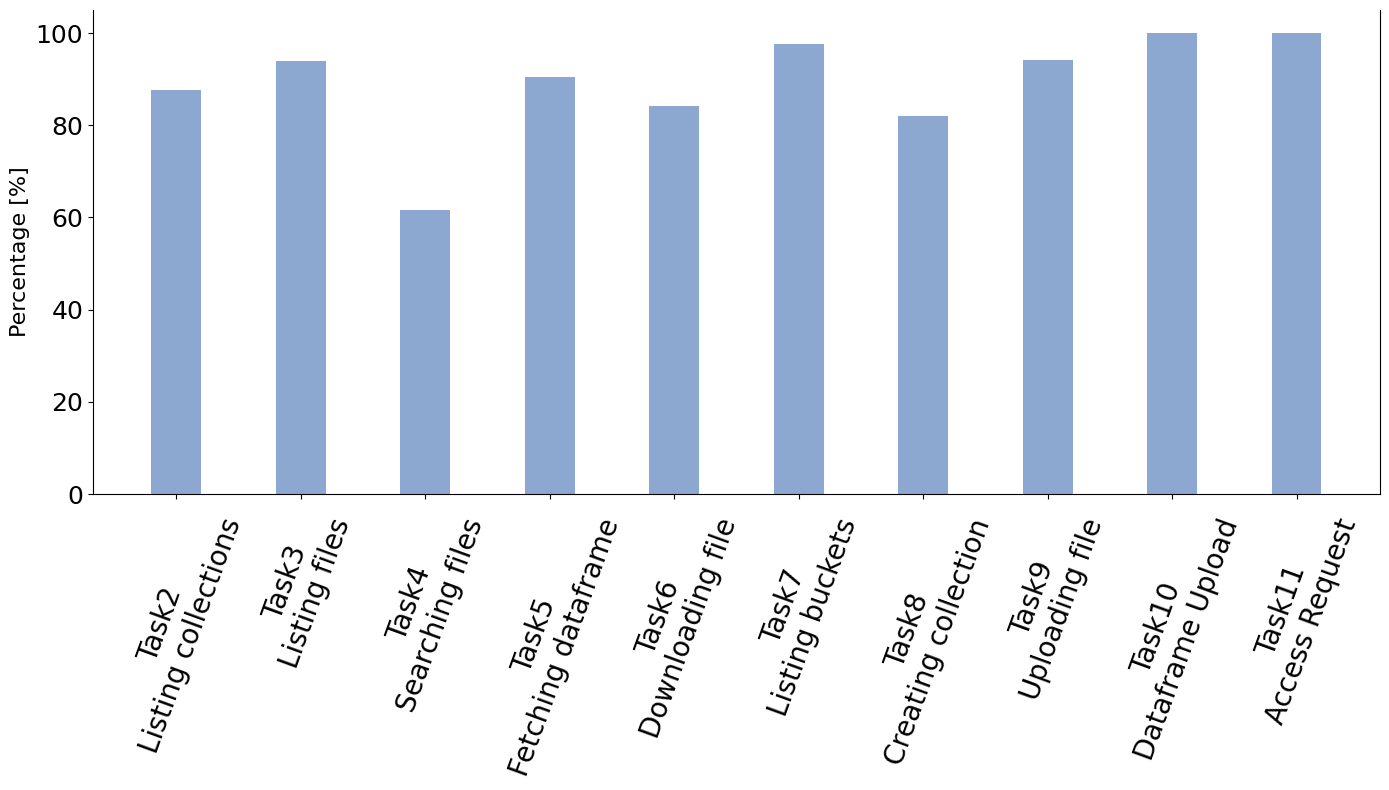} 
\caption{Rate of correct responses per task. Tasks 0 and 1 are excluded since they did not assess critical prototype functionalities.}
\label{fig:correct_answers_percentage}
\end{figure}

We also evaluated the proportion of correct answers given by the participants per task to confirm whether the time spent per task is enough to evaluate the difficulty of a task (see Figure~\ref{fig:correct_answers_percentage}). The results indicate a high level of response accuracy, with a large proportion of correct answers per task among the participants. Tasks 0 and 1 are excluded from this analysis because they did not assess critical functionalities. Task 4 showed the highest rate of inaccurate responses. By combining completion time and answer accuracy, Task 4 can be identified as the most difficult task, suggesting that participants may have had difficulties understanding either the task description or the functionality itself, despite none of them reporting Task 4 as difficult in the usability responses shown in Figure~\ref{fig:usability_per_task}.

\begin{figure}[h]%
\centering
\includegraphics[width=1\linewidth]{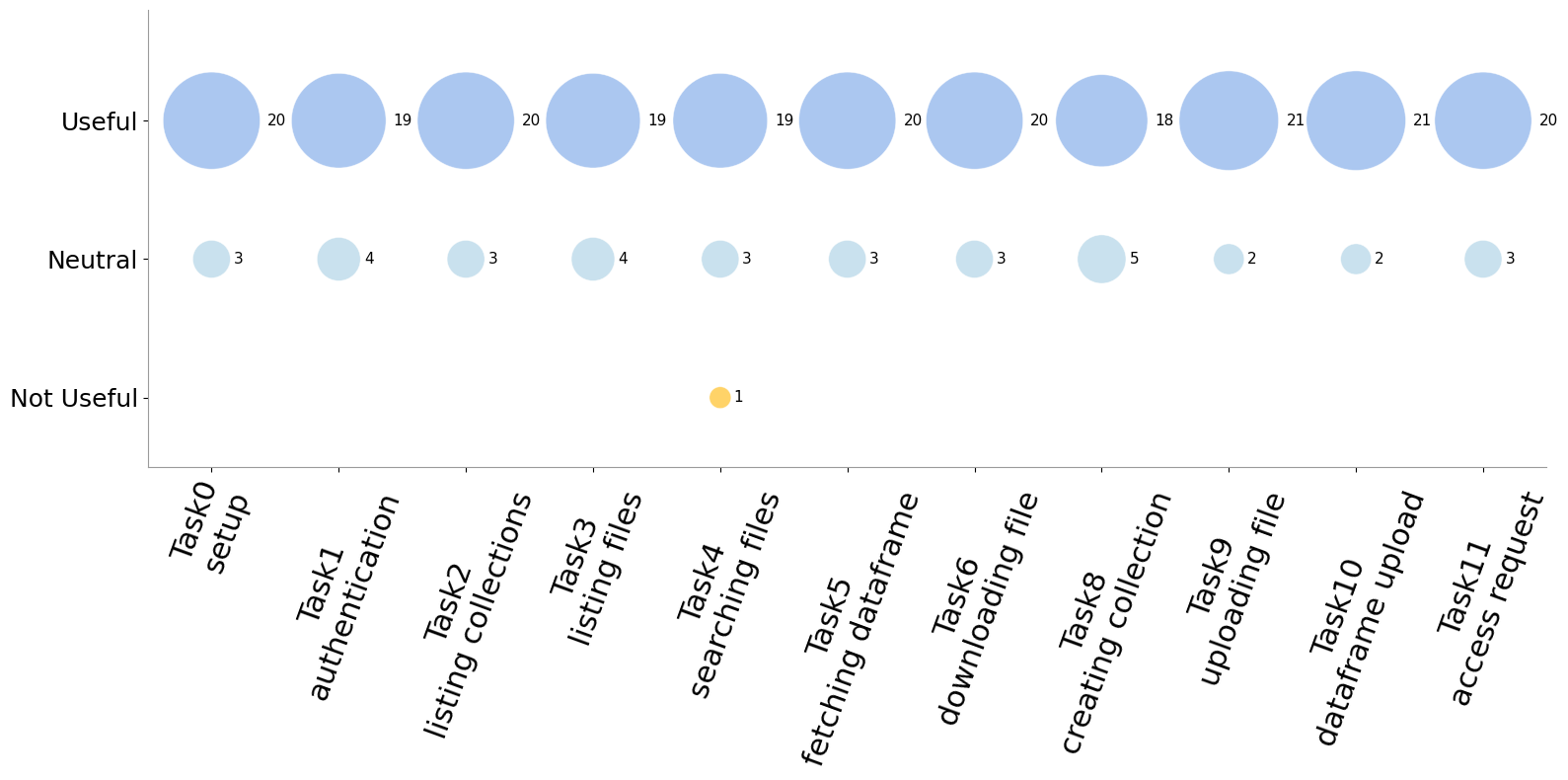} 

\caption[Usefulness responses for each functionality.]{Usefulness responses for each functionality. The usefulness of each task was rated in the final feedback form using three options: ``Useful'', ``Neutral'', and ``Not Useful''. Most participants rated the tasks and their associated functionalities as ``Useful'', while only a few rated them as ``Neutral'' or ``Not Useful''.}

\label{fig:usefulness_per_task}
\end{figure}

To conclude our usability study analysis, we processed the data collected in the final feedback form to evaluate the usefulness of each task. As illustrated in the Figure~\ref{fig:usefulness_per_task}, the answers indicate that most participants found the functionalities useful. These results can be attributed to the fact that the functionalities tested are fundamental to data management. Given that all participants had prior exposure to data management needs, they were likely already familiar with the importance of the concepts and operations involved. The only response classifying the \texttt{search\_files} functionality as not useful is most likely due to the participants' bad experience with the documentation or the task description itself, since we collected additional user feedback through a text field in the form reporting difficulty in understanding the advanced filter usage in the documentation.

\section{Conclusion}

A key takeaway of this paper is that up-to-date, accurate, and ready-to-consume data is invaluable for researchers and users who require reliable information to effectively conduct their research or to make good management decisions. Data management systems such as data lakes, data warehouses, or even non-relational databases may be sufficient for most use cases. However, as identified in previous studies, for more complex use cases, especially inter-institutional research groups, a lakehouse architecture seems to be the most appropriate option due to its design paradigm, which combines the best features of all these approaches. This is embodied through its flexible system design, which allows the storage of heterogeneous data as well as highly organised data used by end-user data consumers. This flexibility makes it more adaptable when project requirements change, which may arise as research projects expand, data volumes grow, and users' needs become more complex.

Storing data in a federated manner helps ensure compliance with data regulations that may vary by geographic location. Distributed storage tools and architectures are essential to implementing data federation. Tools such as Hadoop HDFS, Ceph, Swift, MongoDB, and HBase, which support distributed storage and processing, are particularly valuable in this context~\cite{silva2025review}. Although focused on health data management use cases, these tools and solutions can be applied to use cases in a variety of fields. 

The implementation of our data management platform using a data lakehouse architecture should be viewed as a prototype and proof-of-concept to showcase the viability and usefulness of a flexible and scalable platform for heterogeneous data storage and data sharing. Viewed in this context, our prototype implementation has fulfilled this goal as showcased by the executed user study described in this paper. Based on the results obtained from the user study, we can conclude that the prototype was functional, and user feedback indicated that most functionalities were easy to use. Furthermore, participants found the Python and R libraries' functionalities intuitive. However, after analysing the time-based performance results alongside the additional qualitative feedback, it became apparent that the task questionnaire and documentation descriptions were not always clear enough, which negatively affected the participants' experience. Lastly, during the user study, we gathered informal additional comments from participants, from which we learnt that library documentation and the included code examples were sometimes confusing and could be improved for non-technical users.

\section{Limitations and Future Work}

Regarding the prototype's implementation, there remains substantial room for improvement. The data version control mechanism is limited and can be redesigned to implement large data versioning tools (including the tools identified in our review study) rather than relying on the Lakehouse API to perform it internally in the storage layer. The current OpenHealth Lake's implementation is also limited in terms of storage environments. Currently, it supports Google Cloud Storage, Amazon S3, and Apache HDFS. This can be expanded to allow for integration with other low-cost storage environments, such as MinIO\footnote{\url{https://www.min.io/}} and Ceph\footnote{\url{https://ceph.io/en/}}, through Python and native software development kits provided by these systems.

In the current implementation, the file upload mechanism provides the user with an upload request endpoint that requires file metadata as input and returns a direct upload URL. A file record is created in the file index (file catalogue) when a new upload request is made, before returning a direct upload URL to the user. If the user fails to upload the file or uploads it after the URL expires, the file will not be physically stored in the system. However, its record will persist in the catalogue. Consequently, this strategy requires a periodic storage cleaning mechanism that identifies ghost file records in the catalogue that are not present in the storage zone and removes them. If this cleaning mechanism fails to execute, inconsistencies may be introduced into the data lake zone.

To address this limitation, an event-driven routine can be incorporated. Using this approach, an event publisher creates an entry in an upload queue when a new upload is requested, and the user receives the direct upload URL. A consumer service then checks the queue after the URL expiration time (or periodically), verifies whether the file exists in the storage zone, and only then creates the corresponding catalogue entry. 

Another limiting factor is the level of granularity supported by the current user access management system. Currently, access management is implemented using a passport strategy, where a visa is issued for each data collection created in the system. These visas can then be granted or revoked, allowing or restricting user access to specific collections. The limitation of this approach is that once a visa is granted, the user gains access to all files within that collection. A second layer of visas could be implemented to provide file-level access granularity for use cases that require more fine-grained permissions.

\section{Acknowledgments}

We thank INFORM Africa Research Hub for providing the use case for this study. Likewise, we thank all the participants of the usability study and partnering institutions (Centre for Epidemic Response and Innovation, Instituto Tecnológico de Aeronáutica, INFORM Africa, and the DS-I consortium) for facilitating the presentation of the data lakehouse prototype during the recruitment process.\\
\textbf{INFORM Africa Research Study Group for D-SI Africa Consortium  -- } Akros: Christina Riley, Anna Winters. Centre for the AIDS Programme of Research in South Africa (CAPRISA): Vivek Naranbhai, Johan van der Molen, Salim Abdool Karim. Consortium for Advanced Research Training in Africa (CARTA): Kennedy Otwombe. Institute of Human Virology Nigeria: Alash’le Abimiku, Sophia Osawe, James Onyemata, Patrick Dakum, Fati Murtala-Ibrahim, Nifarta Andrew, Aminu Musa, Tolulope Adenekan, Kenneth Ewerem, Victoria Etuk. Stellenbosch University/ Centre for Epidemic Response and Innovation (CERI): Tulio de Oliveira, Cheryl Baxter, EduanWilkinson, Houriiyah Tegally, Jenicca Poongavanan, Michelle Parker, Danilo Silva, Joicymara S. Xavier. University of Maryland Baltimore: Kristen A. Stafford, Manhattan Charurat, NataliaBlanco, Meagan Fitzpatrick, Mohammad M. Sajadi. University of Port Harcourt. Olanrewaju Lawal.  Villanova University: Chenfeng Xiong, Weiyu Luo, Xin Wu.

\section{Funding}

This work was supported by the National Institutes of Health (NIH) through the INFORM Africa project and the Institute of Human Virology Nigeria [U54 TW012041]; and through eLwazi Open Data Science Platform and Coordinating Center [U2CEB032224].

\section{Competing Interests}

None declared


\bibliographystyle{unsrtnat}
\bibliography{references}


\begin{appendices}

\section{Figures}

This appendix presents supplementary figures that provide additional context for the prototype platform implementation and the user study that evaluated the platform's usability and usefulness.

\clearpage

\begin{figure*}[!ht]%
\centering
\includegraphics[width=0.8\linewidth]{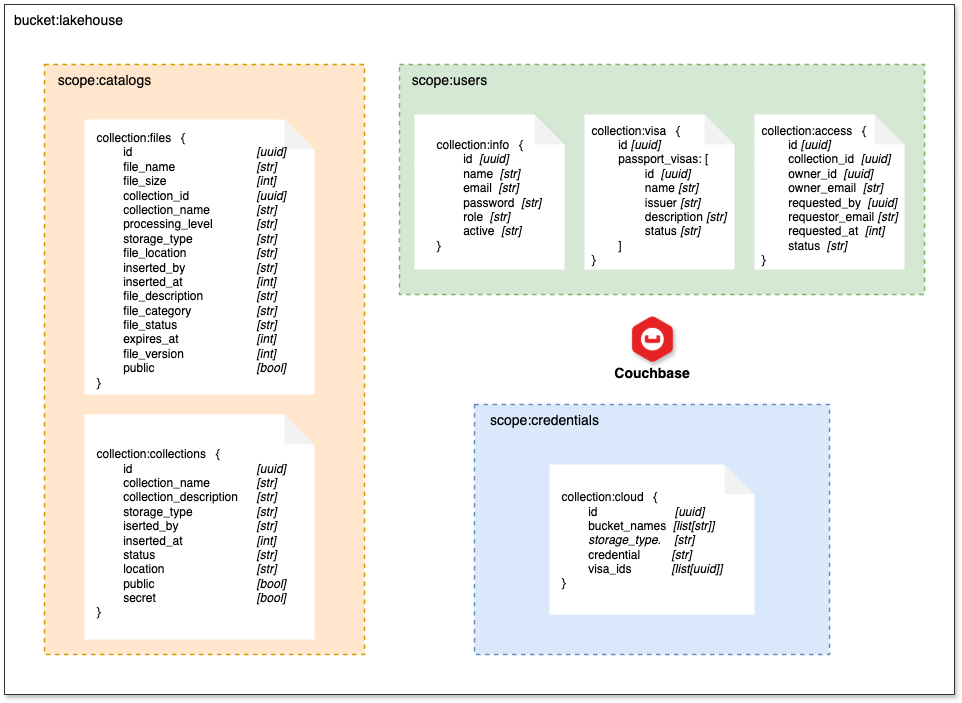} 

\caption{Lakehouse application database model.}

\label{fig:database_model}
\end{figure*}

\begin{figure*}[!h]%
\centering
\includegraphics[width=0.8\linewidth]{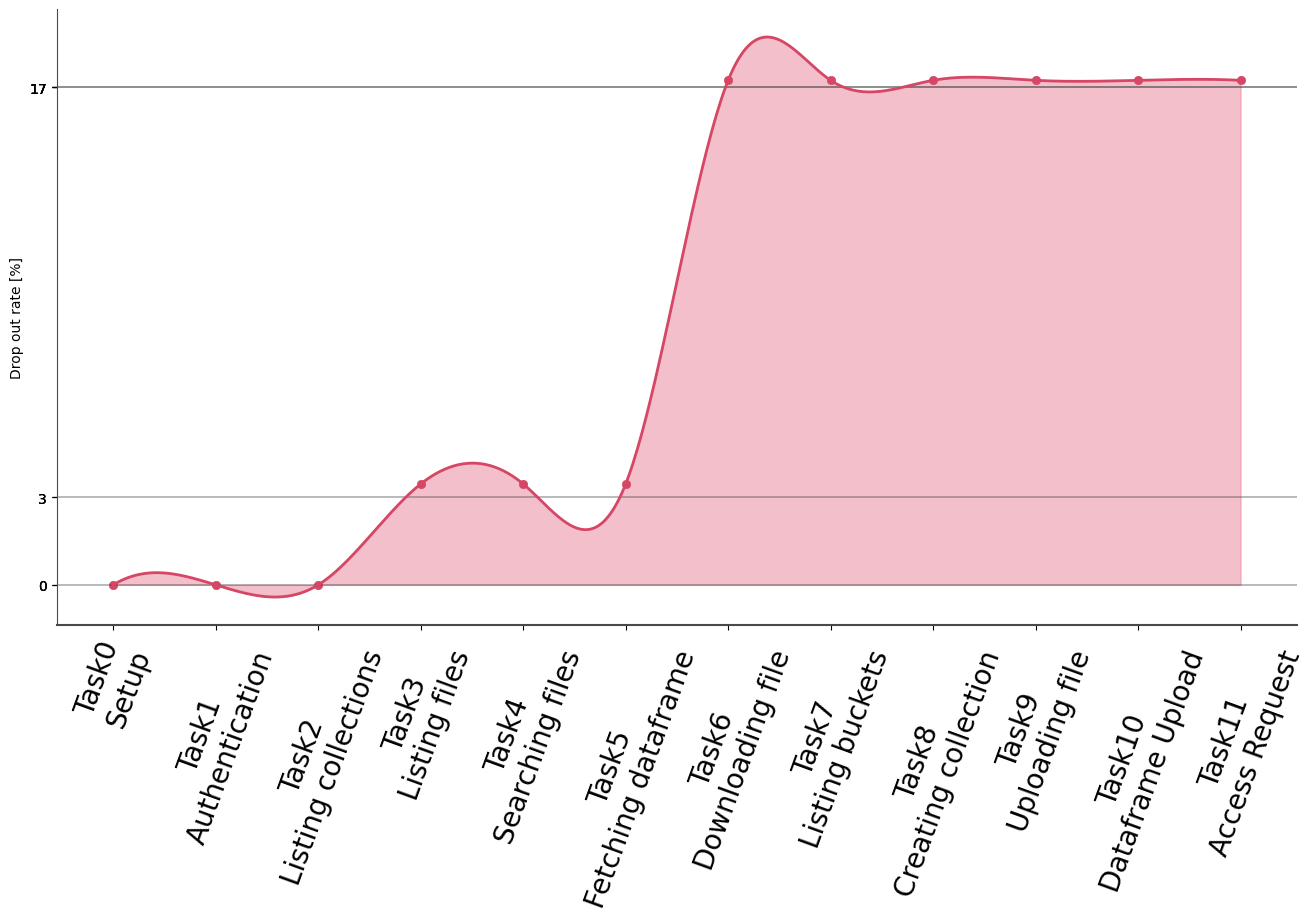} 

\caption{User study dropout rate.}

\label{fig:dropout_rate}
\end{figure*}





\end{appendices}

\end{document}